**High quality epitaxial thin films and exchange bias of antiferromagnetic Dirac semimetal FeSn**

Durga Khadka,[1,#] T. R. Thapaliya,[1,#] Jiajia Wen,[2] Ryan F. Need,[3] and S. X. Huang[1,*]

[1]Department of Physics, University of Miami, Coral Gables, Florida, 33146, USA

[2]Stanford Institute for Materials and Energy Sciences, SLAC National Accelerator Laboratory, Menlo Park, California 94025, USA

[3]Department of Materials Science and Engineering, University of Florida, Gainesville, Florida, 32611, USA

[#]Contributed equally.

[*]sxhuang@miami.edu

Abstract:

FeSn is a topological semimetal (TSM) and kagome antiferromagnet (AFM) composed of alternating $Fe_3Sn$ kagome planes and honeycomb Sn planes. This unique structure gives rise to exotic features in the band structures such as the coexistence of Dirac cones and flatbands near the Fermi level, fully spin-polarized 2D surface Dirac fermions, and the ability to open a large gap in the Dirac cone by reorienting the Néel vector. In this work, we report the synthesis of high quality epitaxial (0001) FeSn films by magnetron sputtering. Using FeSn/Py heterostructures, we show a large exchange bias effect that reaches an exchange field of 220 Oe at 5 K, providing unambiguous evidence of antiferromagnetism and strong interlayer exchange coupling in our films. Field cycling studies show steep initial training effects, highlighting the complex magnetic interactions and anisotropy. Importantly, our work provides a simple, alternative means to fabricate FeSn films and heterostructures, making it easier to explore the topological physics of AFM TSMs and develop FeSn-based spintronics.

While ferromagnets (FM) have historically played the central role in computing technologies, a series of seminal papers have recently highlighted the power and potential of antiferromagnets (AFM) to enable THz-speed memory and logic devices in the post-Moore era[1-8]. One of the key hurdles for AFM spintronics is the development of AFMs in which the spin order (i.e. Néel vector) can be electrically read and reoriented. Antiferromagnetic topological semimetals (TSM),[9,10] are excellent candidates in this regard. In these materials, large spin-orbit coupling and band inversion create Dirac-like states near the Fermi level in which the charge and spin degrees of freedom are intrinsically coupled. At the same time, AFM-TSMs retain all the desirable characteristics of a topologically trivial AFM (e.g., near zero magnetization), thereby providing both a platform to study the interplay between band topology and magnetism, and an opportunity for electrical switching of the Néel vector in AFM spintronics.[11,12]

Among the few known AFM-TSMs, the kagome magnets $Mn_3Sn$[13-15] and FeSn[16,17] are prominent examples. FeSn and $Mn_3Sn$ both feature layered crystal structures containing 2D kagome lattice planes (i.e., $Fe_3Sn$ or $Mn_3Sn$ layers).[16] Each kagome plane is composed of Fe (Mn) atoms that form corner-sharing triangles and Sn atoms that occupy the center of the hexagons [Fig. 1(a)]. In FeSn, with space group P6/mmm and lattice constants $a = 5.298$ Å and $c = 4.448$ Å, $Fe_3Sn$ kagome layers are separated by a honeycomb Sn layer [Fig. 1(b)].[18] The large separation between $Fe_3Sn$ layers created by the Sn spacers isolates the kagome planes and pushes the behavior of FeSn towards the 2D limit of a single kagome plane.[16] In each $Fe_3Sn$ layer, Fe atoms carry ordered magnetic moment of 1.85 $\mu_B$/Fe and order ferromagnetically.[18] However, neighboring $Fe_3Sn$ layers are antiferromagnetically coupled with Néel temperature $T_N \approx 370$ K.[16,18]

Recent angle-resolved photoemission measurements on FeSn single crystals have demonstrated some remarkable features in the bandstructure[16,17]: 1) coexistence of Dirac fermions and flatbands within 0.5 eV of the Fermi level; 2) fully spin-polarized 2D surface Dirac fermions, and 3) nondispersive excitations near $E_F$. In addition, DFT calculations on FeSn show that reorientation of the Néel vector from *ab*-plane to *c*-axis breaks $S_{2z}$ symmetry and drives a transition from gapless Dirac fermions to gapped massive Dirac fermions opening up a large energy gap of 70 meV.[17] These results highlight the interplay between topological band structure, magnetism, and electrical transport in FeSn as well as its potential for spintronic applications. For example, the fully spin-polarized 2D surface Dirac fermions could be exploited for zero-field spin orbit torque switching[19] of magnetization with low energy dissipation in an exchange-bias device between FeSn and adjacent ferromagnet. Similarly, the large gap opening may lead to large magnetoresistance and a sensitive means of electrically detecting changes in the Néel vector and the AFM memory devices that use Néel vectors for data bits.

For these applications in emerging AFM spintronics, thin films are key. Thus far there have been just two reports on the synthesis of FeSn thin films on $SrTiO_3$ and $LaAlO_3$ substrates, both using molecular beam epitaxy (MBE).[20,21] In this work, we report the fabrication of high quality epitaxial (0001) FeSn films by magnetron sputtering. In addition, we combine FeSn with permalloy (Py, Ni/Fe 81/19 wt%) and show large exchange bias and training effects in FeSn/Py heterostructures. These results provide an unambiguous evidence of antiferromagnetism in our FeSn films and demonstrate strong exchange coupling at the interface.

Our films were grown by co-sputtering Fe and Sn targets in a high vacuum magnetron sputtering system with base pressure better than $5 \times 10^{-8}$ torr. The substrate temperature $T_S$ was around 400 ℃. Films were deposited in 5 mTorr of ultra-high pure Ar with a net deposition rate (FeSn) of about 0.8 Å/s. The stoichiometry of FeSn was evaluated from the sputtering rates of Fe and Sn and confirmed by the x-ray diffraction (XRD) phase scan that shows no noticeable impurity phases. FeSn films were grown on (0001) $Al_2O_3$ substrates, which has hexagonal structure and lattice constant $a = 4.758$ Å. However, the lattice constant of FeSn is about 11% larger than that of $Al_2O_3$. To overcome the large lattice mismatch and realize epitaxial growth, we grew a high quality (111) Pt (~ 3 nm) seed layer on the $Al_2O_3$ substrates, followed by the deposition of a (0001) Ru layer. Ru has hexagonal crystal structure with lattice constant $a = 2.706$ Å ($2a = 5.412$ Å), only about 2% mismatch with FeSn. FeSn films were grown on these Ru/Pt buffers, then the samples were cooled down and a protective 2 nm Ru cap layer was deposited at room temperature.

Figure 2(a) shows a XRD 2θ-ω (phase) scan of an 80 nm FeSn film. Strong (000$l$) ($l$ = 2,4,8) diffraction peaks are observed without noticeable impurity peaks. In Fig. 2(b), the (0004) peak of a 29 nm film is accompanied by clear Laue oscillations with period of 0.3º, providing initial indications of high crystalline quality and sharp interfaces. Crystal quality was further examined using high resolution XRD rocking curves (ω scan). The full width half maximum (FWHM) of the rocking curve with a triple axis analyzer is only 0.002º [Fig. 2(c)], which matches the FWHM of the $Al_2O_3$ single crystal substrate. The resolution-limited triple axis FWHM indicates a nearly perfect planar alignment (mosaic) between (0001) FeSn planes and the film plane, likely related to the layered structure of FeSn. With conventional double axis analyzer, the rocking curve appears as a sharp peak sitting on top of a broad peak with FWHM around 0.5º, indicating an appreciable $d$-spacing spreading that may originate from relaxation of strain and/or small compositional variations. The in-plane φ scans [Fig. 2(d)] at $Al_2O_3(11\bar{2}6)$, Pt(200), Ru($11\bar{2}2$), and FeSn($10\bar{1}1$) show discrete diffraction peaks, giving the epitaxial relationship ($Al_2O_3[10\bar{1}0] \parallel Pt[110] \parallel Ru[2\bar{1}\bar{1}0] \parallel FeSn[2\bar{1}\bar{1}0]$) between the substrate, seed layers, and FeSn.

Next, we present electric transport and magnetization results from (0001) FeSn films. The electrical contacts were made on rectangular samples (~ 4 mm × 2 mm) by wire bonding 25 μm Al wires. Figure 3(a) shows the temperature-dependent resistivity $\rho_{xx}$ of a Pt(3nm)/Ru(3nm)/FeSn(80nm) film. The room temperature resistivity was 249 μΩ·cm, close to the reported value in bulk single crystals.[16,18] As temperature decreases, $\rho_{xx}$ decreases until reaching a residual value of 34 μΩ·cm at 5 K. The residual resistivity ratio (RRR), defined as $\rho_{xx}$(300 K)/ $\rho_{xx}$(5 K), is about 7.3, indicating the low defect density and again the excellent crystalline quality of our films. Magnetization measurements with field in the $ab$-plane show a nearly linear diamagnetic signal dominated by the contribution of the $Al_2O_3$ substrate [inset of Fig. 3(a)]. A small ferromagnetic moment is also observed and may be due to slight off-stoichiometry of FeSn, though the contribution is quite small. For example, if the ferromagnetism is due to a small excess of Fe and the formation of Fe clusters ($M$ = 1746 emu/cc), then the measured magnetization value of <10 emu/cc corresponds to a ferromagnetic phase fraction below 0.5%. We also see a clear deviation from linearity in the Hall resistivity ($\rho_H$) at low field as shown in Fig. 3(b). This may be related to the small ferromagnetic component of our films. However, we note that a similar deviation from $\rho_H$ linearity was observed in FeSn films grown by MBE and attributed to multi-band transport.[20] Nevertheless, $\rho_H$ reaches about 0.1 μΩ·cm at $H = 10$ kOe, which is again close to the value in bulk single crystals.[16]

To confirm the FeSn films were AFM ordered and study exchange coupling with FMs, we fabricated a FeSn/Py heterostructure and studied its exchange bias effect.[22,23] Specifically, we measured the anisotropic magnetoresistance (AMR) of Py (Ni/Fe 81/19 wt%) and extracted the coercivities as a function of temperature and loop number. Figure 4(a) shows resistance ($R$) as a function of in-plane magnetic field, $H$, for $I\|H$ and $I\perp H$ at $T = 16$ K, where $I$ is the electric current. The $R$-$H$ loops show typical AMR behavior which gives $R(I\|H) > R(I\perp H)$. The peaks/valleys in the $R$-$H$ curves identify the coercivities ($H_C$) of field-descending branch ($H_{C-}$) and field-ascending branch ($H_{C+}$). Note that temperature fluctuations do not significantly affect our measurements in the temperature range studied here, as evidenced by the presence of closed loops above the saturation field [e.g., Fig. 4(a)]. Importantly, $H_{C-}$ is about -930 Oe while $H_{C+}$ is only 510 Oe, demonstrating a pronounced exchange bias effect. The observation of large exchange bias in FeSn/Py provides an unambiguous evidence of antiferromagnetism in our FeSn films and demonstrates that strong exchange interactions exist between the Py spins and the uncompensated spins of the FeSn (0001) plane at the interface.

Figure 4(b) shows $H_C$ ($H_{C+}$ and $|H_{C-}|$) as a function of temperature for FeSn/Py (at each temperature, the sample was field cooled at $H = 3000$ Oe from $T = 350$ K). Single layer Py films were also measured as a control. As expected, a large increase of $H_C$ is observed upon cooling in the FeSn/Py film, but not in $H_C$ of Py. The $H_C$ of a single 10 nm Py film is only about 1 Oe from $T = 300$K to $T = 25$ K, and increases to 50 Oe at $T = 5$K. In sharp contrast, $|H_{C-}|$ ($H_{C+}$) of FeSn/Py is 15 Oe (14 Oe) at $T = 330$ K, and increases to 913 Oe (495 Oe) which is nearly 3 orders larger than the $H_C$ of Py at $T = 27$ K. The enhancement of $H_C$ at 330 K (our instrumental limit) indicates that the Néel temperature ($T_N$) of FeSn films is above 330 K, and thus reasonably close to the value for bulk FeSn (370 K). $|H_{C-}|$ and $H_{C+}$ of FeSn/Py overlaps at $T \sim 300$ K (difference within 1 Oe), suggesting the blocking temperature ($T_B$) is around 300 K. Note that $T_B$ is notably lower than $T_N$. This may be due to the fact that Fe$_3$Sn and Sn terminations coexist in the interface in which Sn layer attenuates the exchange coupling.

The exchange bias field $H_{EB}$, defined as ($H_{C-} + H_{C+}$)/2, emerges at $T \sim 300$K and increases to -220 Oe at $T = 5$ K [Fig. 4(c)]. Moreover, we observe significant training effect of $H_{EB}$ as a function of loop number $n$ [Fig. 4(d)]. $H_{EB}(n)/H_{EB}(n=1)$ decreases sharply to 0.25 ($T = 149$ K) and 0.43 ($T = 5$ K) at $n = 2$. The large exchange bias field and steep training effect from loop 1 to loop 2 suggest multiple equivalent easy magnetic axes[24], and is consistent with spins lying in the kagome plane with 6-fold rotational symmetry. The multiple easy axes enable initial non-collinear alignments of antiferromagnetic spins, which relax into collinear alignments after the loop 1.[24] Interestingly, $H_{EB}(n=2)/H_{EB}(n=1)$ increases sharply from 200 K to 60 K and remains roughly unchanged between 60 K and 5 K [inset of Fig. 4(d)]. This is likely due to the relaxation of non-collinear alignments during the field cooling process. Further cycling ($n > 2$) results in a continuous decrease in the exchange bias field, but at a markedly slower rate which reflects thermal depinning mechanism.[25] The large exchange bias field and training effect may be exploited to manipulate domain structures of FeSn and develop AFM spintronic devices.

In summary, we report the fabrication of high quality epitaxial films of antiferromagnetic Dirac semimetal FeSn by magnetron sputtering. Significantly, we observed strong interfacial exchange coupling in a FeSn/Py heterostructure evidenced by a large exchange bias field of 220 Oe at 5 K and coercivity fields ~3 orders of magnitude larger than control Py films at 25 K. The availability of high quality FeSn films and the observation of exchange coupling between FeSn and Py open up opportunities to explore topological antiferromagnetic spintronics through thin film engineering.

J.W. acknowledges the support by the Department of Energy, Office of Science, Basic Energy Sciences, Materials Sciences and Engineering Division, under contract DE-AC02-76SF00515.

Data Availability

The data that supports the findings of this study are available within the article.

Figure captions

**Fig. 1.** (a) $Fe_3Sn$ layer with Fe kagome lattice. (b) Crystal structure of FeSn with stacking of $Fe_3Sn$ and Sn layers.

**Fig. 2.** (a) XRD $2\theta$ scan of an 80 nm (0001) FeSn film. (b): XRD $2\theta$ scan of a 29 nm FeSn film showing (0004) peak accompanied by clear Laue oscillations. (c) Triple axis HR-XRD $\omega$ scan (rocking curve) of FeSn (0004) peak. Inset of (c): Double axis HR-XRD $\omega$ scan. (d) In-plane $\varphi$ scans of FeSn, Ru, Pt, and $Al_2O_3$.

**Fig. 3.** (a) Resistivity as a function of temperature for an 80 nm FeSn film. Inset: Magnetic hysteresis loop (field in the *ab*-plane). (b) Hall resistivity as a function of field for a 60 nm FeSn film at $T = 300$ K.

**Fig. 4.** (a) Resistance as a function of in-plane field for FeSn(30nm)/Py(10nm) film with $I \parallel H$ (solid squares) and $I \perp H$ (open circles), respectively. (b) Negative (squares) and positive (circles) coercivities as a function of temperature for FeSn/Py (blue and red symbols) and 10 nm Py (black symbols), respectively. (c) Exchange bias field ($H_{EB}$) of FeSn/Py as a function of temperature. (d) $H_{EB}(n)/H_{EB}(n=1)$ as a function of loop number *n* at $T = 5$ K (solid star), and $T = 149$ K (open star), respectively. Inset: $H_{EB}(n=2)/H_{EB}(n=1)$ as a function of temperature.

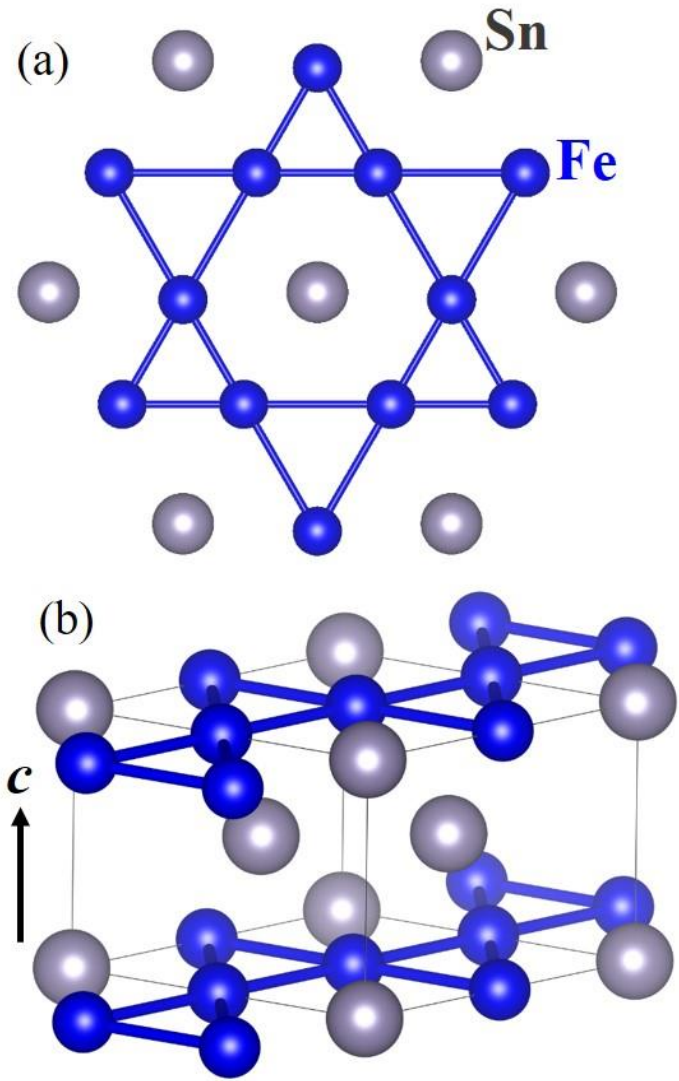

**Fig. 1.** (a) Fe$_3$Sn layer with Fe kagome lattice. (b) Crystal structure of FeSn with stacking of Fe$_3$Sn and Sn layers.

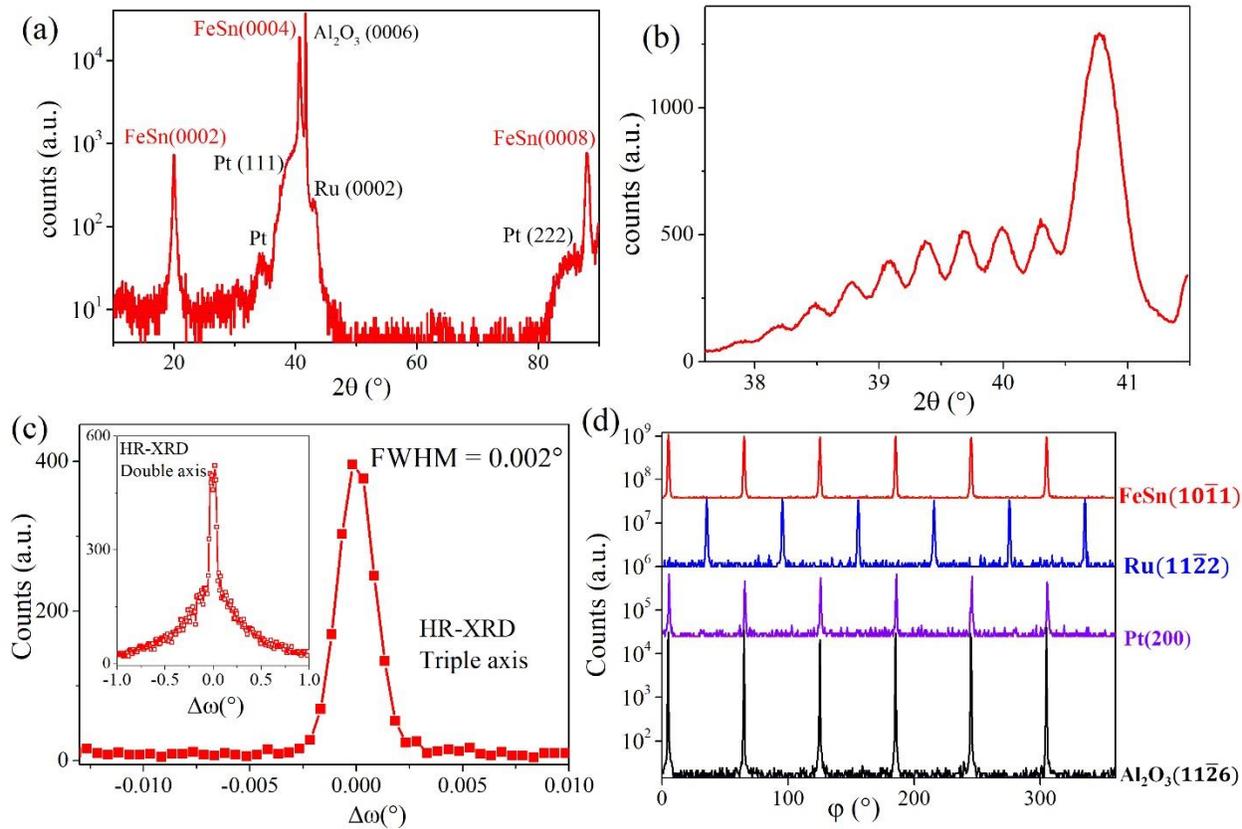

**Fig. 2.** (a) XRD 2θ scan of an 80 nm (0001) FeSn film. (b): XRD 2θ scan of a 29 nm FeSn film showing (0004) peak accompanied by clear Laue oscillations. (c) Triple axis HR-XRD ω scan (rocking curve) of FeSn (0004) peak. Inset of (c): Double axis HR-XRD ω scan. (d) In-plane φ scans of FeSn, Ru, Pt, and $Al_2O_3$.

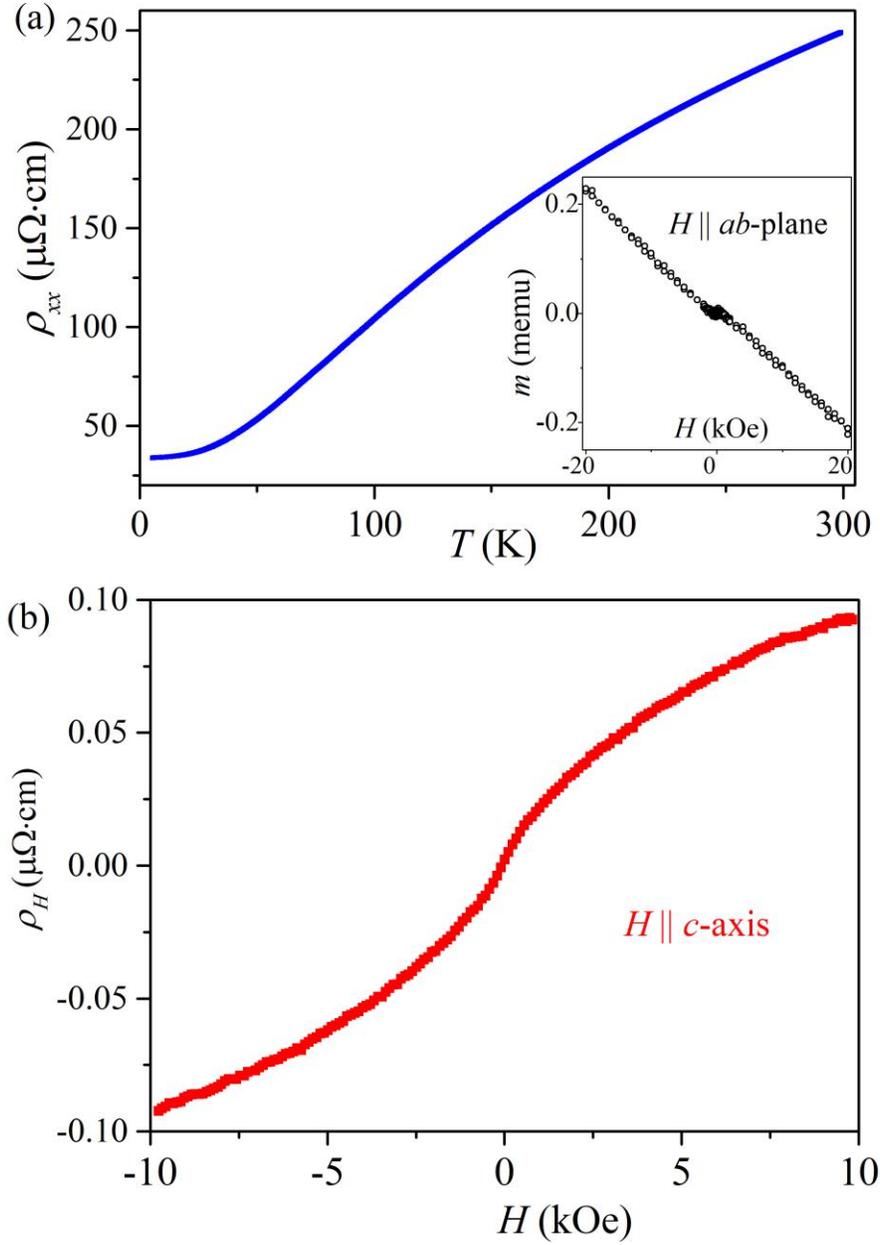

**Fig. 3.** (a) Resistivity as a function of temperature for an 80 nm FeSn film. Inset: Magnetic hysteresis loop (field in the *ab*-plane). (b) Hall resistivity as a function of field for a 60 nm FeSn film at $T = 300$ K.

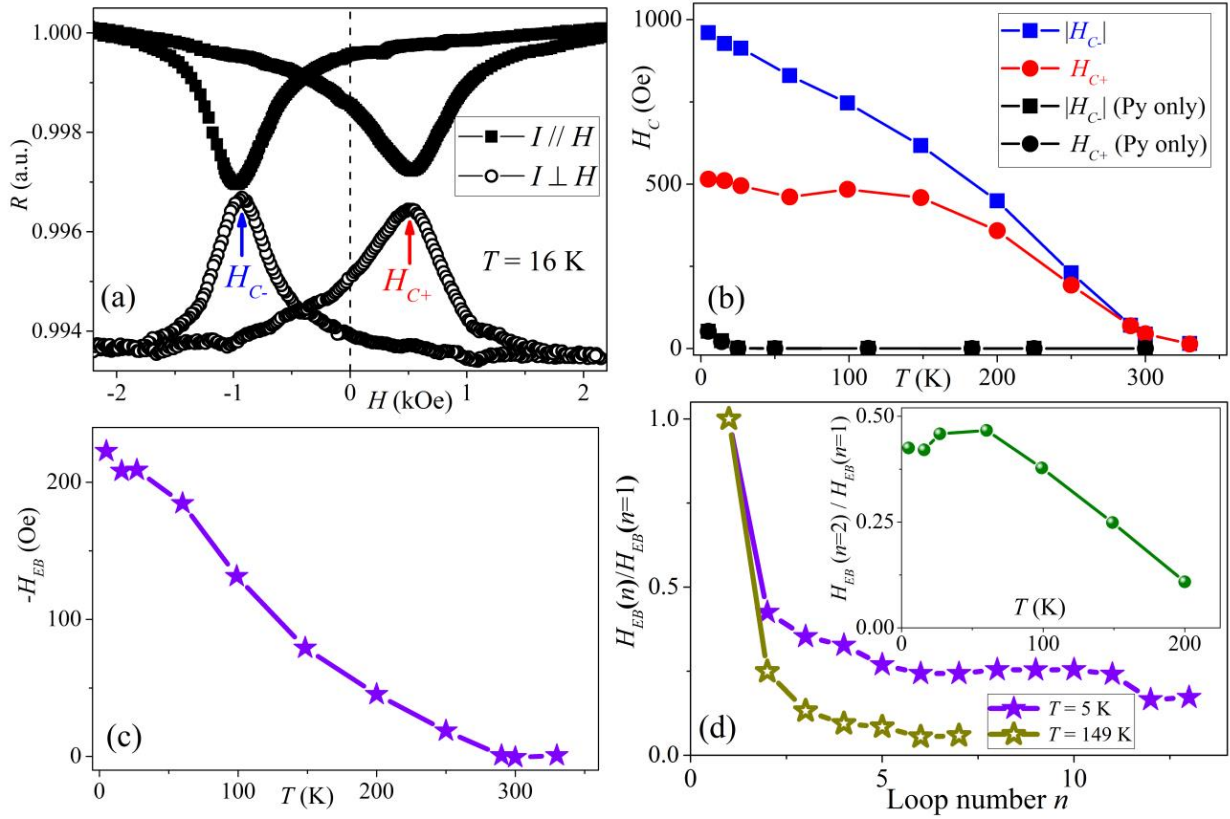

**Fig. 4.** (a) Resistance as a function of in-plane field for FeSn(30nm)/Py(10nm) film with $I \,//\, H$ (solid squares) and $I \perp H$ (open circles), respectively. (b) Negative (squares) and positive (circles) coercivities as a function of temperature for FeSn/Py (blue and red symbols) and 10 nm Py (black symbols), respectively. (c) Exchange bias field ($H_{EB}$) of FeSn/Py as a function of temperature. (d) $H_{EB}(n)/H_{EB}(n=1)$ as a function of loop number $n$ at $T = 5$ K (solid star), and $T = 149$ K (open star), respectively. Inset: $H_{EB}(n=2)/H_{EB}(n=1)$ as a function of temperature.